\documentclass[aps,pra,twocolumn,a4paper,english]{revtex4}
\usepackage{lipsum}

\usepackage{enumerate}
\usepackage[pdftex]{graphicx}
\usepackage{amsmath}
\usepackage{array}
\usepackage{amssymb}
\usepackage{bm}
\usepackage{makecell}
\usepackage{multirow}
\usepackage{color}
\usepackage{subfigure}

\usepackage{algorithmic}
\usepackage{algorithm}

\makeatletter
\newcommand{\removelatexerror}{\let\@latex@error\@gobble}
\makeatother

\begin{document}
	\title{Automatic Depth Optimization for Quantum Approximate Optimization Algorithm}
	
	\author{Yu Pan}
	\email{ypan@zju.edu.cn}
	
	\author{Yifan Tong}
	\email{yftong@zju.edu.cn}
	
	\affiliation{State Key Laboratory of Industrial Control Technology, Institute of Cyber-Systems and Control, College of Control Science and Engineering, Zhejiang University, Hangzhou, 310027, China}
	
	\author{Yi Yang}
	\email{yi.yang6@mcgill.ca}
	
	\affiliation{Department of Mathematics and Statistics, McGill University, Montreal, QC H3A 0B9, Canada}
	
	\begin{abstract}
		Quantum Approximate Optimization Algorithm (QAOA) is a hybrid algorithm whose control parameters are classically optimized. In addition to the variational parameters, the right choice of hyperparameter is crucial for improving the performance of any optimization model. Control depth, or the number of variational parameters, is considered as the most important hyperparameter for QAOA. In this paper we investigate the control depth selection with an automatic algorithm based on proximal gradient descent. The performances of the automatic algorithm are demonstrated on $7$-node and $10$-node Max-Cut problems, which show that the control depth can be significantly reduced during the iteration while achieving an sufficient level of optimization accuracy. With theoretical convergence guarantee, the proposed algorithm can be used as an efficient tool for choosing the appropriate control depth as a replacement of random search or empirical rules. Moreover, the reduction of control depth will induce a significant reduction in the number of quantum gates in circuit, which improves the applicability of QAOA on Noisy Intermediate-scale Quantum (NISQ) devices.
	\end{abstract}
	
	\date{\today}
	
	\maketitle
	
	\section{Introduction}
	Quantum Approximate Optimization Algorithm (QAOA) is considered as one of the most promising applications for near-term Noisy Intermediate Scale Quantum (NISQ) \cite{Preskill18} computing device to demonstrate quantum advantage \cite{farhi2014quantum,farhi2015quantum,farhi2019quantum}. The goal of QAOA is to prepare a quantum state that yields an approximate solution to a classical optimization problem. QAOA is a hybrid quantum-classical algorithm, in the sense that the quantum state is generated and measured using quantum hardware while the control parameters are optimized by the classical algorithm to form a closed loop. In the standard QAOA, the quantum state is generated by executing $p$ blocks of noncommuting quantum operations which consist of $2p$ variational control parameters in total. The control parameters can be optimized by gradient-free \cite{farhi2014quantum,McClean_2016,guerreschi2017,verdon2019}, gradient-based \cite{guerreschi2017,WHJR18,Sweke2020,Daniel20} and machine learning methods \cite{otterbach2017,wu2020end,SRLYP20,pmlr-v107-yao20a,WP20}. In this case, the control depth is $2p$, which is a hyperparameter that has a strong influence on the performance of the model.
	
	Although it has been proven that QAOA converges to the optimal solution in the $p\rightarrow\infty$ limit \cite{farhi2014quantum}, an extremely large $p$ is not physically realizable due to the noise effect and limited control capability of NISQ devices. In practice, the control depth has a finite value which is often preselected. However, although there have been studies on the dependence of QAOA performances upon the control or circuit depth, it is still not clear how to determine an optimal depth with respect to any specific problem \cite{niu2019optimizing,ZW20,larkin2020,majumdar2021optimizing,RJT21}. The current research (e.g., \cite{farhi2014quantum,JRW17,hastings2018,WHJR18,BARE20,RJT21}) have found several empirical or analytical rules to select the control depth for certain problems, which cannot be formulated as automated algorithms. In addition, considering that the empirical or analytical selection rules are different for different problems or even different class of instances, a generic selection rule seems unlikely to exist. However, searching for a good control depth by hand-tuning is computationally inefficient. In particular, none of the current works have investigated any automatic algorithm framework for balancing model accuracy and model complexity with an intermediate control depth, which is critical for the robust implementation of QAOA.
	
This paper presents the first attempt to derive a generic and automatic algorithm for optimizing the control depth of QAOA, which is more efficient than random search and more generally applicable than existing empirical or analytical selection rules. Since the control depth is a hyperparameter of the model, the depth optimization can be taken as a model selection problem. Therefore, any model selection method \cite{JWHT13,LUXBURG2011651} can be considered to solve this problem. In this paper, we employ the model selection method with $l_1$ regularization (by additionally minimizing the $l_1$-norm of a parameter vector) \cite{RT96} for mainly two reasons. First, the regularized model can be optimized iteratively, which makes it very efficient. For example, regularization techniques such as LASSO (Least Absolute Shrinkage and Selection Operator) \cite{RT96} and its variants are often used for the automatic model selection. LASSO imposes an $l_1$ regularization on the parameters to be estimated in addition to the objective function, which can effectively shrink the number of parameters of the model during iteration. Second, an $l_1$ regularized optimization problem can be solved by fast algorithms with optimality and convergence guarantee. In particular, the commonly-used Proximal Gradient (PG) descent method can be used to solve a linear and convex problem with a basic convergence rate of $\mathcal O(1/k)$ \cite{Combettes2011} which can be further accelerated by a line search \cite{AM09}, where $k$ is the current iteration step. Since the objective function in QAOA is possibly non-convex \cite{ZW20}, we also have to consider an extension of PG descent algorithm that is compatible with both convex and non-convex problems \cite{HZ15,ijcai2017-462}. Other regularization terms can also be used as candidates if certain assumptions on the correlations between the parameters are made \cite{Zou05}.
	
The most common criterion for model selection is to achieve the balance between model accuracy and complexity \cite{JWHT13,LUXBURG2011651}. It should be noted that for the noise-free case in this paper, the accuracy of the model will be monotonically increasing with the control depth. As a result, a criterion that balances the accuracy and control depth has to be proposed, otherwise the best model would be the unregularized one if accuracy is the sole consideration. Fortunately, the necessity of complexity control naturally arises in the context of QAOA. To be more specific, the objective for QAOA can be stated as optimizing the approximation ratio until a desired value has been reached, which has been proven to be NP-hard with classical algorithms \cite{ZW20}. In other words, we only require QAOA to give an optimization result that is good enough instead of the best result. In this case, the optimization target can be expressed as
	\begin{equation}
		\frac{C^{*}}{C_{\max}}\geq r^{*},\label{intro1}
	\end{equation}
	where $C^{*}$ is the optimized value of the objective function and $C_{\max}$ is the maximum value. $C_{\max}$ can be obtained by brute force methods across different instances \cite{khairy2020learning} or enforcing a frustration-free Hamiltonian whose ground-state energy is known in advance \cite{FWC09}. Here $C^{*}/C_{\max}$ stands for the approximation ratio. According to (\ref{intro1}), the approximation ratio measures the optimization accuracy. An approximate solution $C^{*}$ is smaller than $C_{\max}$, and QAOA intends to produce a solution that is as close to $C_{\max}$ as possible. Finding the minimum depth that satisfies $(\ref{intro1})$ with a good-enough $r^{*}$ naturally confines the control complexity and is crucial for the understanding of the fundamental performance limitations of QAOA \cite{APM20,CRAB21}, which can further contribute to a better design that mitigates the circuit complexity in the practical implementation.
	
	In this paper, we provide a comprehensive study of the performance of PG descent methods for the control depth optimization of QAOA. Numerical results show that the performance of QAOA is heavily affected by the control depth, which can be significantly optimized during the iteration using a proper regularization strength. This paper is organized as follows. A short introduction on QAOA is given in Section \ref{Sec:SQ}. The algorithms for control depth optimization are detailed in Section \ref{Sec:ADO}. In Section \ref{Sec:MA}, the Max-Cut problem and metrics for numerical experiment are defined. Section \ref{Sec:NR} presents the numerical results on 7-node and 10-node examples. Conclusion is summarized in Section \ref{Sec:C}.
	
	\section{Standard QAOA}\label{Sec:SQ}
	The Pauli matrices are defined by
	\begin{equation}
		\sigma_z=\left(
		\begin{array}{cc}
			1& 0 \\
			0& -1
		\end{array}
		\right),\quad \sigma_x=\left(
		\begin{array}{cc}
			0& 1 \\
			1& 0
		\end{array}
		\right).
	\end{equation}
	The two computational basis states of a single qubit are denoted as $|0\rangle=(0\quad 1)^T$ and $|1\rangle=(1\quad 0)^T$. Let $H_o$ be the Hamiltonian that encodes the corresponding optimization problem, and $H_c=\sum_{n=1}^N\sigma_{x}^{(n)}$ be the control Hamiltonian. The optimal solution to the classical problem, which is often a combinatorial problem, is encoded as the ground state of $H_o$. The standard QAOA is executed by alternatively applying the non-commuting $H_o$ and $H_c$ as
	\begin{equation}
		|\psi(\beta,\gamma)\rangle=e^{-\mbox{i}H_c\beta_p}e^{-\mbox{i}H_o\gamma_p} \cdot\cdot\cdot e^{-\mbox{i}H_c\beta_1}e^{-\mbox{i}H_o\gamma_1}|s\rangle.\label{cp}
	\end{equation}
	Here $|s\rangle=|+\rangle_1\cdot\cdot\cdot|+\rangle_N$ is the initial $N$-qubit state that can be easily prepared, with $|+\rangle=1/\sqrt{2}(|0\rangle+|1\rangle)$. The parameters $\beta=(\beta_1,\cdot\cdot\cdot,\beta_p),\gamma=(\gamma_1,\cdot\cdot\cdot,\gamma_p)$ stand for control angles constrained in the region $(-\pi,\pi]$, and $2p$ is the control depth. A negative parameter corresponds to the angle of a backward rotation. The goal of QAOA can be formulated as
	\begin{equation}
		\min_{\beta,\gamma}f(\beta,\gamma)=\min_{\beta,\gamma}\langle\psi(\beta,\gamma)|H_o|\psi(\beta,\gamma)\rangle.\label{pmin}
	\end{equation}
	That is, the goal is to optimize $\beta,\gamma$ such that the generated quantum state $|\psi(\beta,\gamma)\rangle$ approximates the ground state of $H_o$. In QAOA, (\ref{cp}) is implemented using quantum hardware. In practice, each unitary operation defined in (\ref{cp}) is realized as a sequence of quantum gates. Moreover, $f(\beta,\gamma)$ in (\ref{pmin}) is the expectation value of $H_o$ with respect to the state $|\psi(\beta,\gamma)\rangle$, which can be calculated by repeatedly executing (\ref{cp}) and taking the average of the quantum measurement results with respect to $H_o$. In contrast, the optimization algorithm for the control parameters $\beta,\gamma$ can be designed and implemented classically.
	
	\section{Algorithms For Depth Selection}\label{Sec:ADO}
	In this paper, we adopt $l_1$ regularization technique for the automatic reduction of control depth. The regularized model is given by
	\begin{eqnarray}
		&&\min_{\beta,\gamma}f(\beta,\gamma)+g(\beta,\gamma)\nonumber\\
		&=&\min_{\beta,\gamma}\langle\psi(\beta,\gamma)|H_o|\psi(\beta,\gamma)\rangle+\lambda\sum_{j=1}^p(|\beta_j|+|\gamma_j|).\label{QAOAobj}
	\end{eqnarray}
	The effect of the additional regularization term $g(\beta,\gamma)$ is shrinking some of the control parameters to be exactly zero during optimization. Letting $x=(\beta,\gamma)$, (\ref{QAOAobj}) can be written in standard form as
	\begin{equation}
		\min_x\langle\psi(x)|H_o|\psi(x)\rangle+\lambda||x||_1,\label{regx}
	\end{equation}
	where $||\cdot||_1$ denotes the $l_1$-norm of a vector. Note that $f(\cdot)$ is differentiable but possibly non-convex \cite{ZW20}, while the regularization term $g(\cdot)$ is convex but non-differentiable.
	
	\begin{figure}[t]
		\includegraphics[scale=0.5]{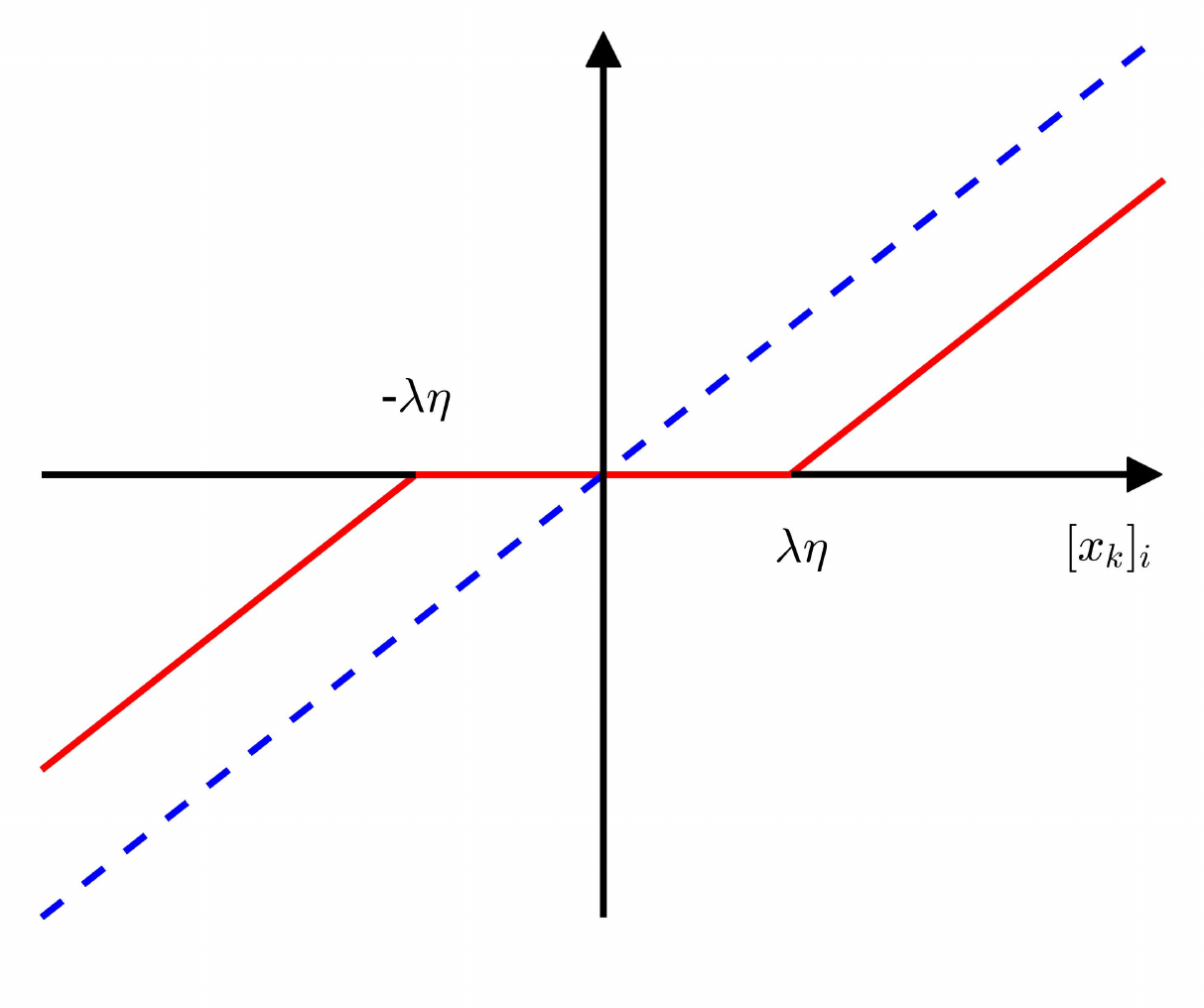}
		\centering
		\caption{The illustration of soft-thresholding operation, with $y$-axis value determines the value of parameter after thresholding. Blue dotted line indicates the reference case without soft-thresholding. During each iteration, the updated control parameter is then shifted towards zero by an amount of $\lambda\eta$ (red line). If the absolute value of the control parameter is below the threshold, then it will be directly penalized to zero (red line).}\label{softhresholding}
	\end{figure}
	
	The PG descent method updates the control parameters as
	\begin{eqnarray}
		&&x_{k+1}=\nonumber\\
		&&\mathop{\arg\min}_{z}f(x_k)+\nabla f(x_k)^T(z-x_k)+\frac{1}{2\eta}||z-x_k||_2^2\nonumber\\
		&&+g(z).\label{pgmin}
	\end{eqnarray}
	The motivation is to minimize the quadratic approximation to $f(\cdot)$ around $x_k$ with $\nabla^2f(\cdot)$ replaced by $\frac{I}{\eta}$, and leave the non-differentiable $g(\cdot)$ alone. Here $\eta$ is the learning rate. For $g(\cdot)=\lambda||\cdot||_1$, the minimization problem (\ref{pgmin}) has the following closed-form solution
	\begin{equation}
		x_{k+1}=S_{\lambda \eta}(x_k-\eta\nabla f(x_k)).\label{pgformula}
	\end{equation}
	Here $S_{\lambda \eta}(x_k)$ is the soft-thresholding operator defined by
	\begin{equation}
		S_{\lambda \eta}({[x_k]}_i)=\left\{
		\begin{array}{rcl}
			&{[x_k]}_i-\lambda\eta,           & {[x_k]}_i>\lambda\eta \\
			&0,    &     -\lambda\eta\leq{[x_k]}_i\leq\lambda\eta\\
			&{[x_k]}_i+\lambda\eta,          & {[x_k]}_i<-\lambda\eta
		\end{array} \right.
	\end{equation}
	where ${[x_k]}_i$ is the $i$-th element of the vector $x_k$, for $i=1,...,2p$. As shown in Fig.~\ref{softhresholding}, once the condition $|{[x_k]}_i|\leq\lambda\eta$ is met, ${[x_k]}_i$ will be penalized to exactly zero, which will remove the corresponding control action and reduce the control depth. The gradient vector $\nabla f(x_k)$ is approximated by
	\begin{equation}
		[\nabla f(x_k)]_i\approx\frac{f({[x_k]}_i+\epsilon)-f({[x_k]}_i-\epsilon)}{2\epsilon},\quad \epsilon>0,\label{fd}
	\end{equation}
	in which $f({[x_k]}_i\pm\epsilon)$ can be estimated using either classical simulation or quantum hardware. It should be noted that the stop criterion is to continue the iterations until a good-enough threshold value has been reached, and thus whether the expectations are obtained by classical simulation or quantum measurement would not matter. Particularly, the gradient descent update did not use any information of the underlying dynamical model. Therefore, the only consequence brought by realistic noise is that the achieved approximation ratio may be lower than the noise-free case, and gradient estimates may be noisy.
	
	Generally speaking, the convergence rate of $\mathcal O(1/k)$ cannot be guaranteed in the absence of the convexity assumption on $f(\cdot)$. Nevertheless, since PG descent is a first-order optimization method just as the stochastic gradient descent, its convergence curve is still expected to be stable. In addition, as can be seen from (\ref{pgformula}), PG descent is simply a reweighting of the conventional gradient descent, which is unlikely to get unstable. Due to these reasons, the stability of the convergence curves can be clearly observed in the numerical results of this paper.
	
	\begin{figure}[!t]
		
		\renewcommand{\algorithmicrequire}{\textbf{Require:}}
		\renewcommand{\algorithmicensure}{\textbf{Initialize:}}
		
		\begin{algorithm}[H]
			\caption{Non-convex APG Algorithm for QAOA}
			\label{APG:implementation}
			\begin{algorithmic}[1]
				\REQUIRE $f(\cdot)$ is $L$-Lipschitz smooth.
				\ENSURE $\eta\in(0,1/L);x_0=x_1;\lambda,tol>0$; a positive integer $q$.
				\FOR{$k = 1, \cdots, K$}
				\STATE $y_k = x_k+\frac{k-1}{k+2}(x_k-x_{k-1});\quad \backslash\backslash$ line search for possible acceleration
				\STATE $F_k=\max_{t\in(\max(1,k-q),...,k)}[f(x_t)+g(x_t)]$;
				\IF{$f(y_k)+g(y_k)\leq F_k$}
				\STATE $\backslash\backslash$ check if $y_k$ is a good extrapolation
				\STATE $v_k=y_k$;
				\ELSE
				\STATE $v_k=x_k$;
				\ENDIF
				\STATE $\nabla f(v_k)$ is calculated using classical simulation or quantum hardware;
				\STATE $x_{k+1}=S_{\lambda \eta}(v_k-\eta\nabla f(v_k));\quad \backslash\backslash$ PG update
				\STATE $f(x_{k+1})$ is calculated using classical simulation or quantum hardware;
				\IF{$|f(x_{k+1})+g(x_{k+1})-F_k|<tol$}
				\RETURN $x_k;\quad \backslash\backslash$ early stopping if convergence has been achieved
				\ENDIF
				\ENDFOR
				\RETURN $x_{K+1}$;
			\end{algorithmic}	
		\end{algorithm}
		
	\end{figure}
	
	We also consider a second algorithm which does not require $f(\cdot)$ to be convex and has a proved convergence rate. This algorithm is an efficient Accelerated Proximal Gradient (APG) descent method proposed in \cite{ijcai2017-462}. We can use this unified approach for both convex and non-convex problems to construct the classical optimization part in QAOA. The details are shown in Algorithm \ref{APG:implementation}. It should be noted that the extrapolation defined by step 2 can accelerate the convergence if $f(\cdot)$ is locally convex. Steps 3-9 are designed to check if the acceleration works. In step 3, choosing $q>0$ allows $y_k$ to occasionally increase the objective function $f(\cdot)$ in order to escape from local minima \cite{LM02,ijcai2017-462}. In particular, $q$ is set to be 2 in the numerical experiments of this paper. Although the only difference between APG and PG appears to be the acceleration part, the convergence rate of APG is shown to be $\mathcal O(1/k)$ for non-convex $f(\cdot)$ \cite{ijcai2017-462}.
	
	The regularization parameter $\lambda$ in (\ref{regx}) is used to control the model complexity. A large $\lambda$ forces most of the elements in $x$ to be zero, while a very small $\lambda$ results in a model that is close to an unregularized one. For this reason, $\lambda$ has to be carefully chosen to encourage a simple control model with sparse $x$ while achieving an sufficient level of optimization accuracy. The common practice is to conduct a grid search to find the optimal $\lambda$. In order to mitigate the resource requirement and running time of the algorithm, the candidate hyperparameters can be examined in the following order
	\begin{equation}
		\lambda_1>\lambda_2>...>\lambda_M.
	\end{equation}
	That is, the algorithms start with a relatively large $\lambda$, which will produce the most sparse control model. Then the procedures are repeated by decreasing $\lambda$, until a predetermined optimization accuracy has been achieved. Since the regularization strength cannot be too large or too small, only a few $\lambda$s have to be tested before an acceptable result has been obtained, irrespective of problem size. In contrast, the number of experiments needed for finding an optimized depth with random search scales as $\mathcal O(p)$. Therefore, the automatic algorithm is very efficient when dealing with large-scale problems, where we have $M\ll p$.
	
	It should be noted that for the noise-free QAOA considered in this paper, the optimization accuracy always tends to increase with the control depth \cite{farhi2014quantum,BVW20}, which means we can always improve the optimization accuracy by decreasing $\lambda$. In that case, the model will become overly complex, or the control depth will become too large. As we have stated, the model complexity can be controlled by selecting the largest $\lambda$ that satisfies the criterion (\ref{intro1}).
	
	\section{Max-Cut and approximation ratio}\label{Sec:MA}
	As a proof-of-principle demonstration, the performance of depth optimization is demonstrated on Max-Cut problem \cite{farhi2015quantum}. Consider an $N$-node non-directed and weighted graph $G=(V,E)$. Max-Cut seeks for a partition of $V$ into two subsets $S_1$ and $S_2$ such that the sum of weights of edges connecting nodes in the two disjoint subsets is maximized. By assigning $Z_i=1,i\in S_1$ and $Z_j=-1,j\in S_2$, the Max-Cut problem can be formulated as a binary optimization
	\begin{equation}
		C_{\max}=\max_{i,j}\frac{1}{2}\sum_{(i,j)\in E}\omega_{ij}(1-Z_iZ_j).\label{cmax}
	\end{equation}
	It is known that finding an approximate solution of binary string $Z^*$ such that
	\begin{equation}
		\frac{C(Z^*)}{C_{\max}}\geq r^*
	\end{equation}
	is NP-hard, with $r^*$ being a minimum ratio \cite{J01,ZW20}. In the QAOA setup, the problem-based Hamiltonian for (\ref{cmax}) is given by
	\begin{equation}
		H_{o^{'}}=\frac{1}{2}\sum_{(i,j)\in E}\omega_{ij}(I-\sigma_{z_i}\sigma_{z_j}).
	\end{equation}
	The approximation ratio for benchmarking the performance of the QAOA can be defined as
	\begin{equation}
		r^{'}=\frac{\max_{\beta,\gamma}\langle\psi(\beta,\gamma)|H_o^{'}|\psi(\beta,\gamma)\rangle}{C_{\max}}.
	\end{equation}
	Particularly, maximization the expectation of $H_{o^{'}}$ is equivalent to the minimization of expectation of the following Hamiltonian
	\begin{equation}
		H_o=\sum_{(i,j)\in E}\omega_{ij}\sigma_{z_i}\sigma_{z_j},
	\end{equation}
	for which the approximation ratio can be defined by
	\begin{equation}
		r=1-\frac{\min_{\beta,\gamma}\langle\psi(\beta,\gamma)|H_o|\psi(\beta,\gamma)\rangle-C_{\min}}{C_{\max}-C_{\min}}\in[0,1],\label{ar}
	\end{equation}
	with $C_{\min}$ being the theoretical best result. According to (\ref{ar}), the approximation ratio $r$ measures the optimization accuracy. In particular, $r=1$ implies that a perfect solution has been obtained by QAOA.
	
	\begin{widetext}
		
		\begin{table}[]
			\caption{Instances of 7-node and 10-node Max-Cut}
			\label{instances}
			\scalebox{0.9}{\begin{tabular}{|l|l|l|l|l|l|l|l|l|l|l|l|l|l|l|}
					\hline
					& Vertex & Weight & Vertex & Weight & Vertex & Weight & Vertex & Weight & Vertex & Weight & Vertex & Weight & Vertex & Weight \\ \hline
					\multirow{2}{*}{7-Node Max-Cut} & $Z_0$ $Z_4$& 0.73 & $Z_0$ $Z_6$ & 0.50 & $Z_1$ $Z_5$  & 0.36   & $Z_2$ $Z_6$  & 0.58   & $Z_3$ $Z_6$  & 0.43   &        &        &        &        \\ \cline{2-15}
					& $Z_0$ $Z_5$& 0.33 & $Z_1$ $Z_4$  & 0.69 & $Z_2$ $Z_5$ & 0.88   & $Z_3$ $Z_5$  & 0.67    &        &        &        &        &        &        \\ \hline
					\multirow{2}{*}{10-Node Max-Cut} & $Z_0$ $Z_1$& 0.21& $Z_0$ $Z_6$  & 0.67 & $Z_1$ $Z_5$  & 0.34   & $Z_1$ $Z_9$  & 0.89  & $Z_2$ $Z_9$  & 0.92   & $Z_3$ $Z_8$  & 0.77   & $Z_4$ $Z_9$  & 0.68\\ \cline{2-15}
					& $Z_0$ $Z_5$& 0.41& $Z_0$ $Z_8$  & 0.82   & $Z_1$ $Z_6$  & 0.77   & $Z_2$ $Z_6$  & 0.45& $Z_3$ $Z_7$  & 0.81   & $Z_4$ $Z_5$  & 0.35   & $Z_8$ $Z_9$  & 0.15 \\ \hline
			\end{tabular}}
		\end{table}
		
	\end{widetext}
	
	\section{Numerical Results}\label{Sec:NR}
	Two randomly generated instances of Max-Cut for the numerical illustration are given in table \ref{instances}. The 7-node instance has 9 weighted edges, and the 10-node instance has 14 weighted edges. {\color{blue} The code and data for the experiments are available in supplemental material \cite{supplement}}.
	
	\subsection{7-node Max-Cut}
	\begin{figure}[t]
		\includegraphics[scale=0.6]{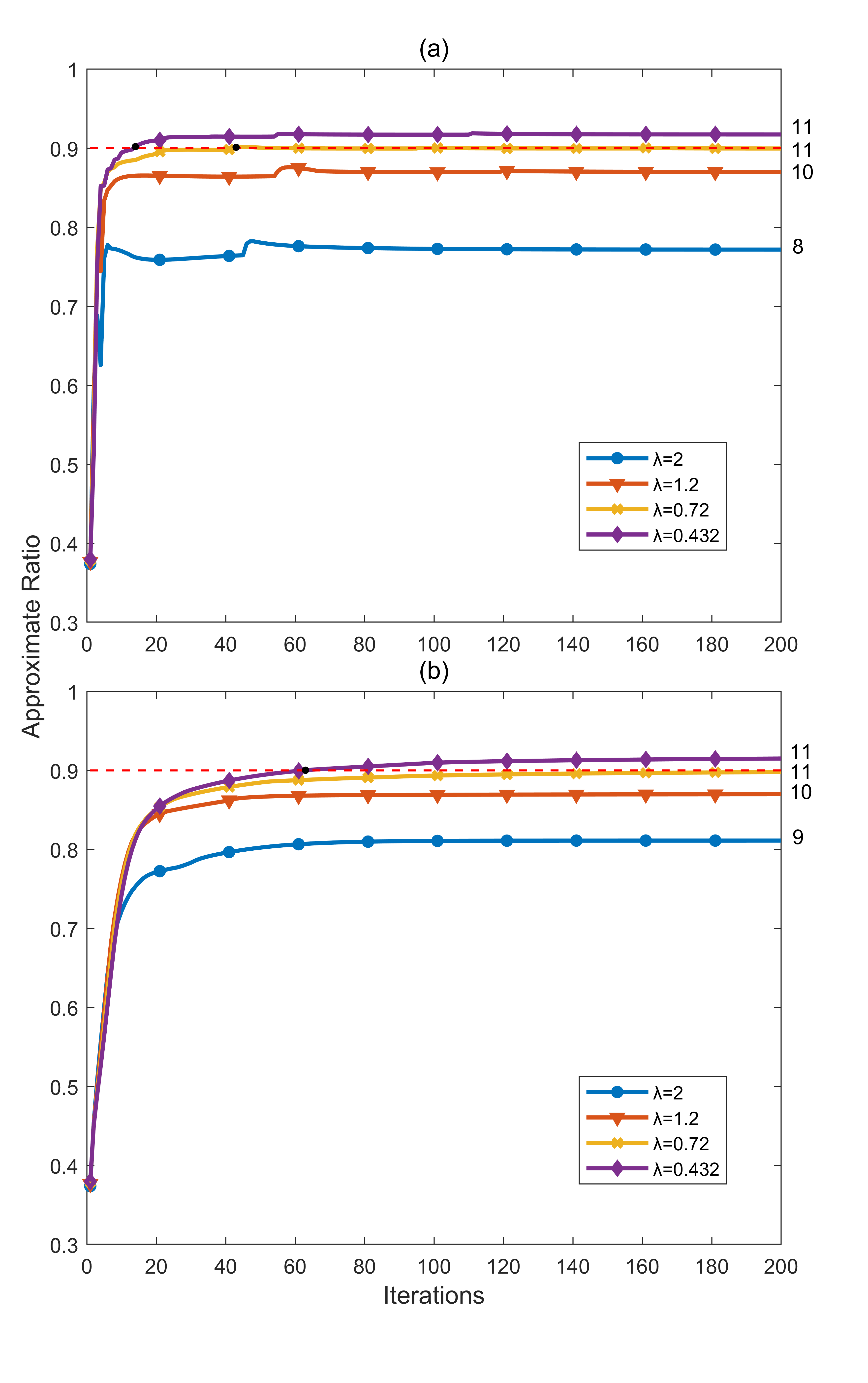}
		\centering
		\caption{
			\label{7-pgapg}
			Approximation ratios versus number of iterations for different $\lambda$. (a) APG algorithm. (b) PG algorithm. The black dots mark for the first time the approximation ratios have reached the minimum acceptable value 0.9. The numbers on the right of the figure refer to the final depths after 200 iterations.}
		
	\end{figure}
	
	\begin{figure}[t]
		\includegraphics[scale=0.07]{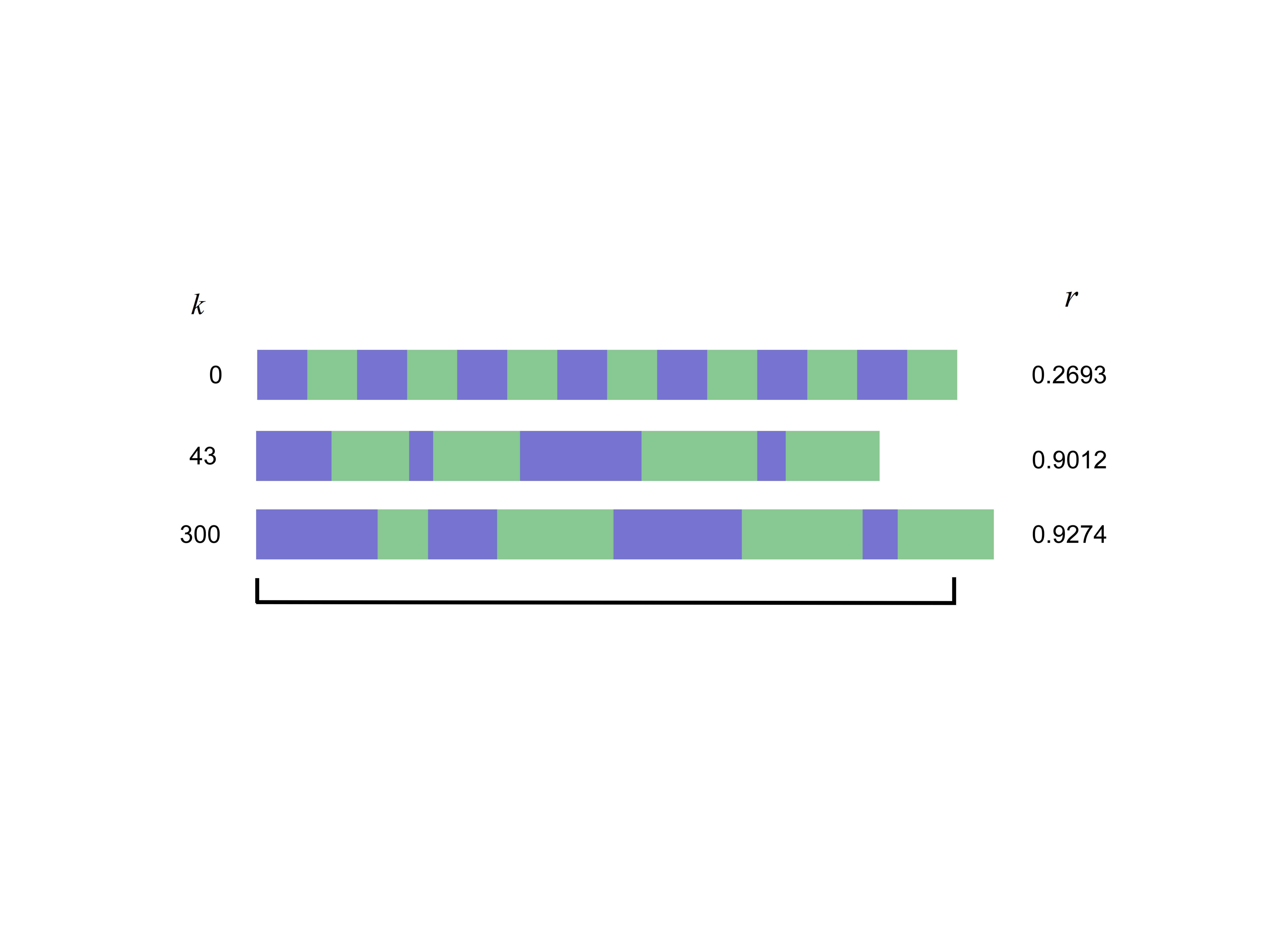}
		\centering
		\caption{
			\label{apgc}The optimized control sequence with $\lambda=0.72$. The first 43 iterations are calculated using the regularized model and APG algorithm. After the approximation ratio has reached 0.9, the regularization term has been removed. Conventional gradient descent is used to continue the optimization process until 300 iterations, with a fixed control depth. The length of each operation is the absolute value of the parameter.}	
	\end{figure}
	
	First, we demonstrate the performance of depth optimization on the 7-node Max-Cut problem. We let $\eta=0.006$. The elements of $\beta$ and $\gamma$ are all initialized as $0.3$, with an initial control depth of $14$. The minimum acceptable approximation ratio is set to be $0.9$, which is drawn in red dashed line in Fig.~\ref{7-pgapg}. The regularization parameter has four values $\{2,1.2,0.72,0.432\}$, which are tested in decreasing order. Note that the effective range of regularization parameter can be estimated based on the range of the objective function and the initial values of the control parameters, as the regularization term has to be comparable with the objective function for the optimization to be successful.
	
	As can be seen in Fig.~\ref{7-pgapg}, the depth selection results are consistent for APG and PG with the same $\lambda$. More specifically, the final control depths are basically the same with the same $\lambda$ after 200 iterations. Also, it has been shown that if $\lambda$ is too large, the algorithms cannot reach the minimum acceptable ratio. The final accuracy is mainly determined by the control depth. That is, the approximation ratios achieved for different algorithms are very close under the same depth, which highlights the importance of depth optimization.
	
	In particular, acceleration in convergence can be observed in Fig.~\ref{7-pgapg}(a). APG converges in less than 50 iterations, while PG may need more than 100 iterations to converge. APG achieves the minimum acceptable ratio in about 40 iterations with $\lambda=0.72$, and the control depth has been shrunk to 11 at that point. Note that the removal of one control parameter may reduce the number of required control operations by two. For example, if $\beta_j$ in the control sequence
	\begin{equation}
		H_o\stackrel{\gamma_j}{\longrightarrow}H_c\stackrel{\beta_j}{\longrightarrow}H_o\stackrel{\gamma_{j+1}}{\longrightarrow}\label{Hdef}
	\end{equation}
	has been penalized to 0, then the two control operations generated by $H_o$ will be combined into one operation, with a duration of $\gamma_j+\gamma_{j+1}$. As a result, the reduction of control depth from $14$ to $11$ implies a more significant reduction in the number of control operations. As shown in Fig.~\ref{apgc}, although the final control depth is $11$, the number of required control operations is only $8$. In this case, the actual percentage of reduction in control operations is $6/14$ that exceeds $40\%$. Since each unitary control operation is realized as a composition of quantum gates, this reduction in control depth will induce a significant reduction in the number of quantum gates, or the depth of quantum circuit, for practical implementation of QAOA.
	
	\begin{figure}[t]
		\includegraphics[scale=0.42]{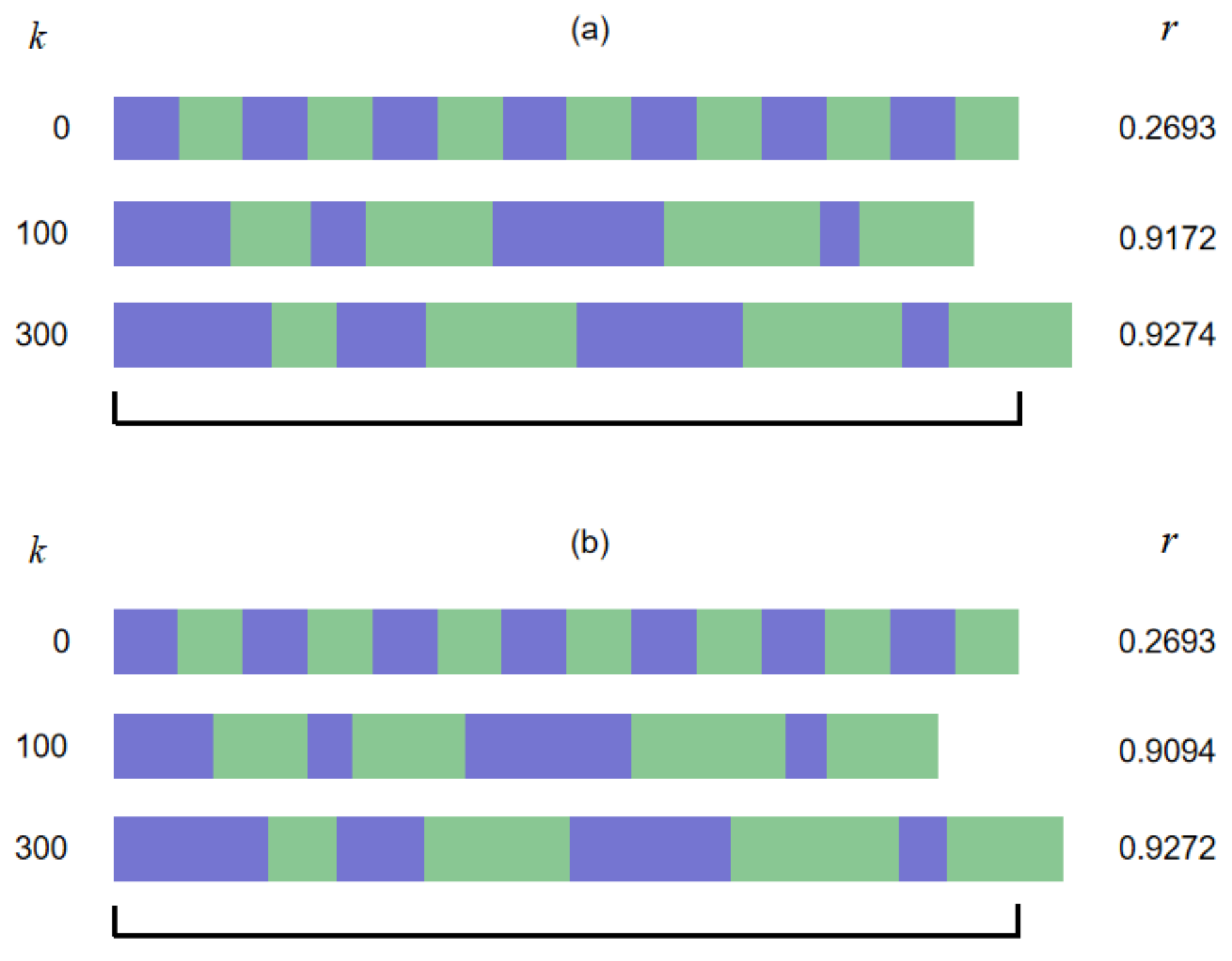}
		\centering
		\caption{
			\label{pgapgc}
			The optimized control sequences for (a) APG algorithm and (b) PG algorithm with $\lambda=0.432$. The first 100 iterations are calculated using the regularized model, and after that the regularization term has been removed. Conventional gradient descent is used to continue the optimization process until 300 iterations, with a fixed control depth.}
		
	\end{figure}
	
	\begin{figure}[t]
		\includegraphics[scale=0.6]{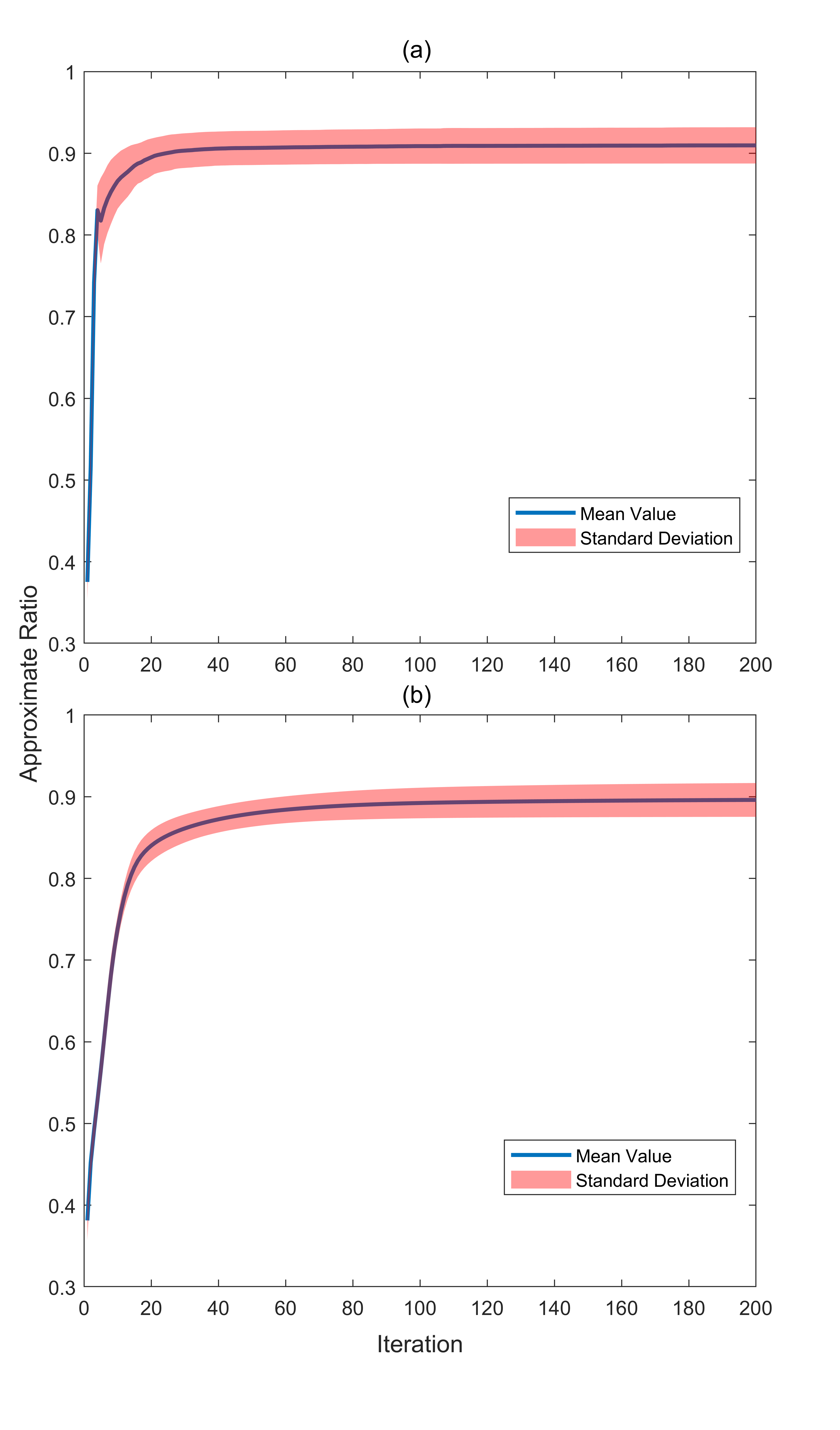}
		\centering
		\caption{
			\label{7-random}
			Mean and standard derivation of the approximation ratios are plotted versus number of iterations for (a) APG algorithm and (b) PG algorithm. The initial values for the 14 control parameters are independently sampled from a uniform distribution over an interval $[0.27,0.33]$.}
		
	\end{figure}
	
	The standard practice for model selection is to use the regularized model to solve for the best hyperparameter, and then continue the optimization process without the regularization term to achieve the best accuracy. We have followed this practice and the optimized control sequences are depicted in Fig.~\ref{apgc} and \ref{pgapgc}. In Fig.~\ref{pgapgc}, the APG and PG algorithms are iterated for 100 times, and then the control depth is fixed for further optimization. It should be noted that the three experiments have achieved basically the same approximation ratio after 300 iterations, which is around 0.9274. This may be because the selected control depth is the same for all three experiments. Started with different values of parameters after the regularization, the ultimate accuracy is still determined by the control depth. In this case, this phenomenon can be taken as evidence to show the importance of optimizing the control depth. Next, we conduct another experiment to show that the performances of APG and PG are not sensitive to small changes in the initial parameters. To do this, the mean and standard derivation of the convergence curves of approximation ratios are calculated with respect to randomly sampled initial values over the interval $[0.27,0.33]$. As shown in Fig.~\ref{7-random}, 100 samples are drawn from a uniform distribution for APG and PG, respectively. The numerical results have confirmed that the convergence of both algorithms are stable against random initialization of parameters.
	
	\begin{figure}[t]
		\includegraphics[scale=0.6]{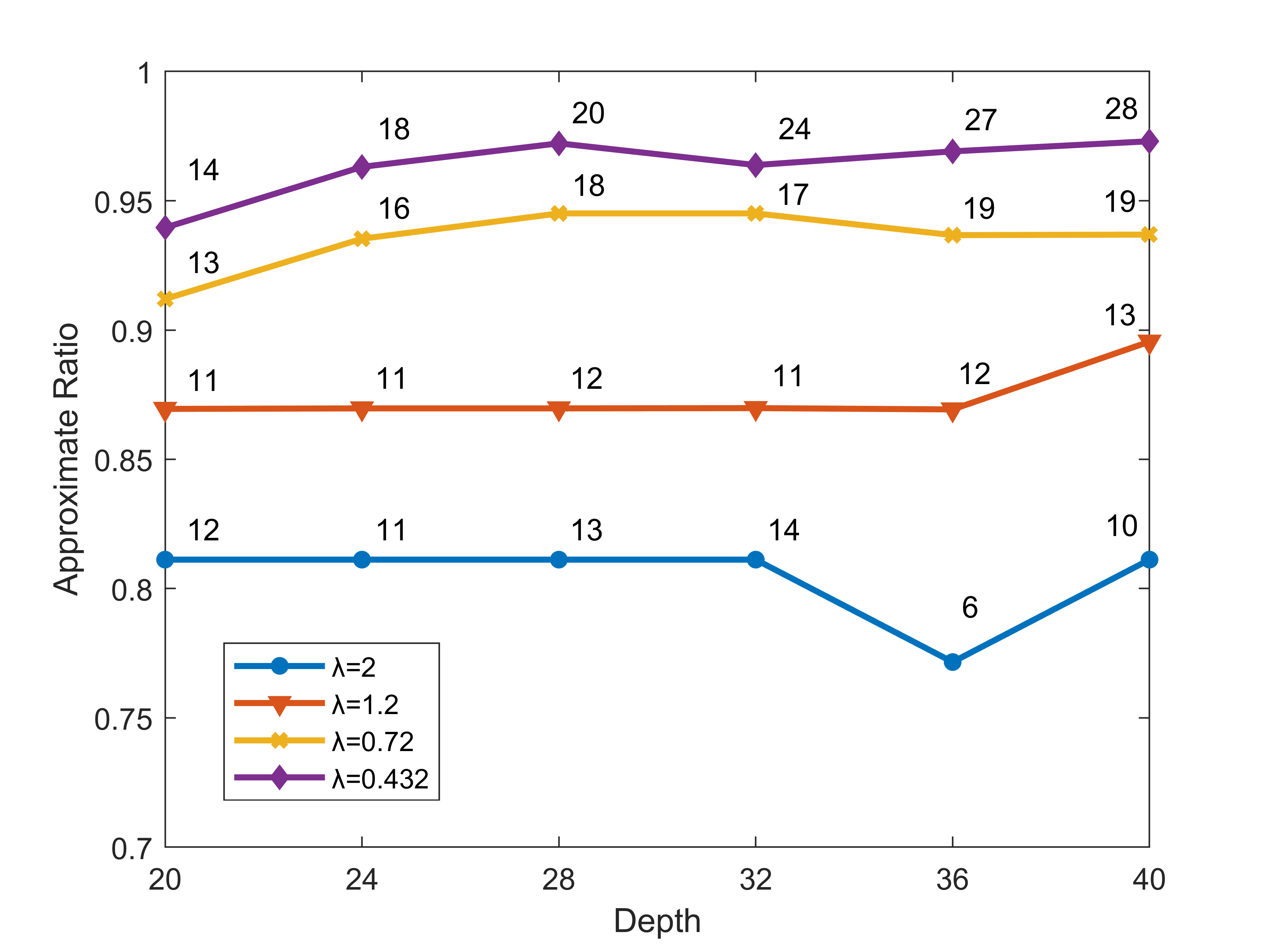}
		\centering
		\caption{
			\label{dep}
			The approximation ratios versus initial control depth for different $\lambda$. The approximation ratio for each initial depth is calculated after 200 iterations of PG algorithm. The numbers along the curve are the final depths after the optimization.}
	\end{figure}
	
	In Fig.~\ref{dep}, we test the performance of depth optimization by varying the initial control depth, from 20 to 40. For $\lambda=2$ and $1.2$, the regularization term dominates and the optimized control depths are all similar for different initial values. Even if a large initial value is chosen, the control depth will be automatically reduced to a small value during the iteration, otherwise the regularization term will become too large to optimize. In this case, both the optimized depth and approximation ratio are steady. For $\lambda=0.72$ and $0.432$, there is a more subtle trade-off between the regularization term and the objective function. In particular, the control depths are allowed to increase slowly to promote the optimization accuracy. This observation is consistent with the existing conclusion that the optimization accuracy will monotonically increase with the control depth for noise-free QAOA. Specifically, the selected control depth doubled as we increase its initial value from $20$ to $40$ for $\lambda=0.432$. In this case, the regularization strength is too small to penalize the parameters.

	\begin{figure}[!htp]
		\includegraphics[scale=0.6]{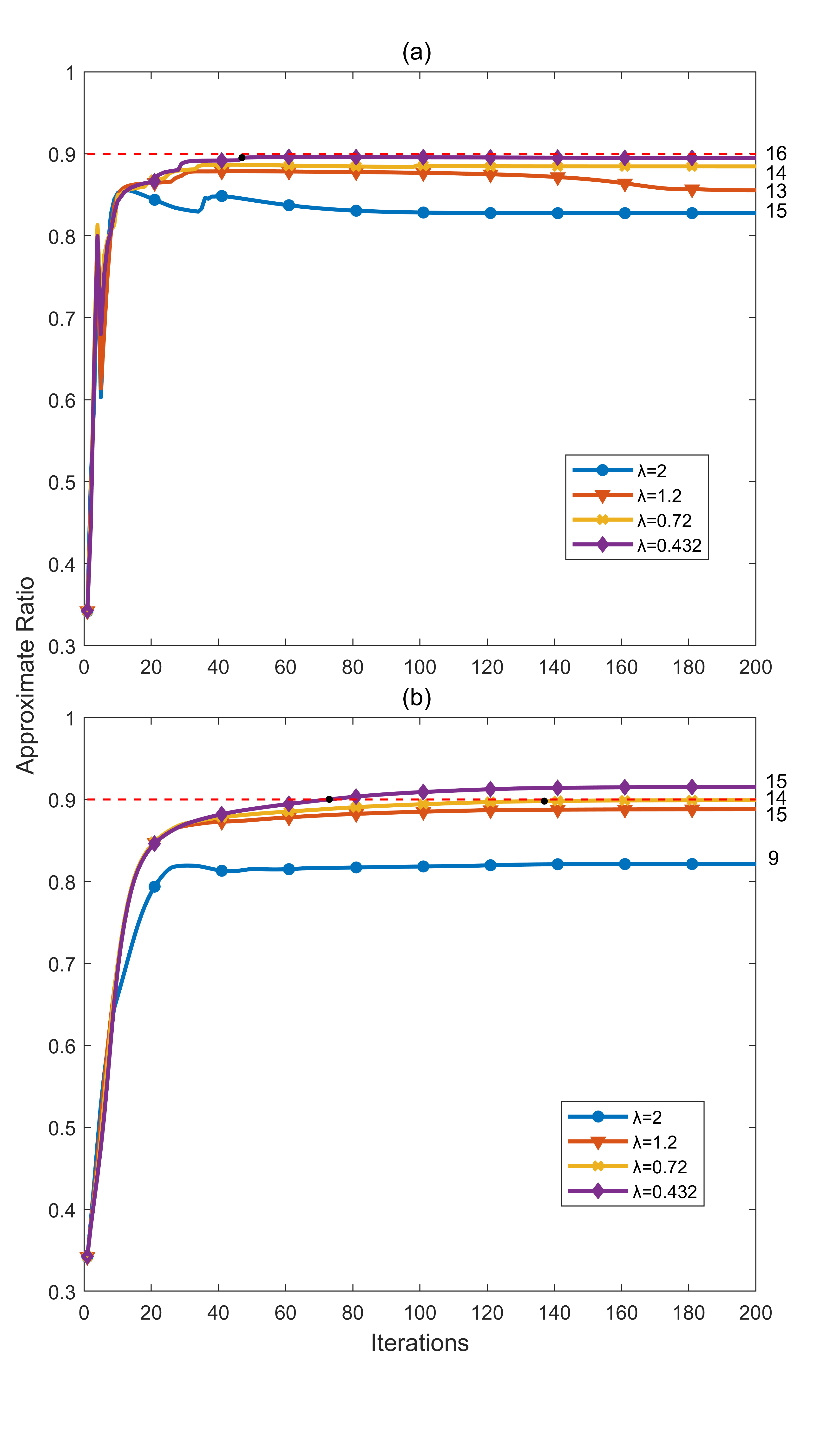}
		\centering
		\caption{
			\label{10-pgapg}
			Approximation ratios versus number of iterations for different $\lambda$. (a) APG algorithm. (b) PG algorithm. The black dots mark for the first time the approximation ratios have reached the minimum acceptable value 0.9. The numbers on the right of the figure are the final depths after 200 iterations.}
		
	\end{figure}
	
	\begin{table}
		\caption{The results are obtained when the approximation ratios reach 0.9 and after 200 iterations, respectively.}
		\label{unc}
		\centering
		\begin{tabular}{|cc|c|c|c|}\hline\multicolumn{2}{|c|}{\multirow{2}{*}{Initial Depth}}               & \multirow{2}{*}{$\lambda = 0$} & \multirow{2}{*}{$\lambda = 0.432$} & \multirow{2}{*}{$\lambda = 0.72$} \\\multicolumn{2}{|c|}{}                                             &                                &                                    &                                   \\ \hline\multicolumn{1}{|c|}{\multirow{6}{*}{20}} & Length($r = 0.9$)      & 5.36382                        & 5.21552                            & 4.9451                            \\ \cline{2-5} \multicolumn{1}{|c|}{}                    & Depth($r = 0.9$)       & 20                             & 19                                 & 15                                \\ \cline{2-5} \multicolumn{1}{|c|}{}                    & Iterations($r = 0.9$)  & 32                             & 27                                 & 46                                \\ \cline{2-5} \multicolumn{1}{|c|}{}                    & Length(200 iterations) & 6.39597                        & 5.11906                            & 4.99829                           \\ \cline{2-5} \multicolumn{1}{|c|}{}                    & Depth(200 iterations)  & 20                             & 14                                 & 13                                \\ \cline{2-5} \multicolumn{1}{|c|}{}                    & $r$(200 iterations)    & 0.9753                         & 0.9396                             & 0.9120                            \\ \hline\multicolumn{1}{|c|}{\multirow{6}{*}{28}} & Length($r = 0.9$)      & 8.95274                        & 7.02808                            & 6.71423                           \\ \cline{2-5} \multicolumn{1}{|c|}{}                    & Depth($r = 0.9$)       & 28                             & 28                                 & 26                                \\ \cline{2-5} \multicolumn{1}{|c|}{}                    & Iterations($r = 0.9$)  & 33                             & 25                                 & 30                                \\ \cline{2-5} \multicolumn{1}{|c|}{}                    & Length(200 iterations) & 9.13553                        & 6.78664                            & 6.51269                           \\ \cline{2-5} \multicolumn{1}{|c|}{}                    & Depth(200 iterations)  & 28                             & 20                                 & 18                                \\ \cline{2-5} \multicolumn{1}{|c|}{}                    & $r$(200 iterations)    & 0.9882                         & 0.9722                             & 0.9451                            \\ \hline\end{tabular}
	\end{table}
	
	We conduct numerical experiments with initial depths 20, 28 and compare the results of regularized and unregularized cases. The detailed results are shown in Table \ref{unc}. In this paper, the performance is measured by the model complexity when approximation ratio reaches a desired threshold value. As can be seen from Table \ref{unc}, the control length, depth and iterations for the regularized models to reach 0.9 are always smaller than the unregularized model, which means that the unregularized model is always the worst and has much room to improve. Even after 200 iterations, the regularized model with $\lambda=0.432$ still has similar accuracy as the unregularized one, while the control length and depth are significantly reduced under the regularization.
	
	\subsection{10-node Max-Cut}
	For the 10-node instance, the initial control depth is set to be 20. The optimization landscape of 10-node instance is more complex than that of 7-node instance, and thus we choose a smaller learning rate $\eta=0.0035$ to stabilize the convergence. The initial values for the control parameters are still $0.3$. The convergence curves for different $\lambda$ are plotted in Fig.~\ref{10-pgapg}. It can be observed that the performance of APG is unstable at initial stage. In contrast, the convergence of PG is always stable during the iteration. At the initial stage, APG seeks for an acceleration by exploring the optimization landscape with extrapolations. Since bad extrapolations are allowed ($q=2$) in the experiments, the convergence curves can be severely perturbed at initial stage that may cause the objective function to fall into bad local minima. Moreover, the ultimate approximation ratio achieved with APG is slightly worse than PG. This may suggest one has to be cautious with APG in high-dimensional setting due to its indeterministic behaviour in acceleration, although it has a proved convergence rate of $\mathcal O(1/k)$.

	\section{Conclusion}\label{Sec:C}
	In this paper, we have proposed an automatic and fast algorithm for depth optimization of QAOA based on proximal gradient descent. On one hand, this algorithm can be used as an efficient tool to study the ultimate performances and limitations of QAOA by optimizing the hyperparameter. On the other hand, the automatic algorithm can be directly integrated with quantum hardware to accelerate the optimization process and reduce the control complexity. In particular, the proper control depth can be found with $M$ experiments, where $M$ is the number of candidate regularization parameters which is small and irrespective of problem size. This results in a significant reduction in the number of experiments for determining the best control depth and approximation ratio, as the number of experiments for finding the best control depth by random search scales with $\mathcal O(p)$, which could be extremely time-consuming for large-scale problems.
	
	There are two interesting directions for future work. Firstly, due to the inevitable noise in NISQ devices, the optimal control depth could be naturally constrained. As the errors induced by the noise accumulate over time and increase with the control depth, the performance of QAOA may start to decline after the control depth reaches an optimal value. This effect can also be studied by adding noise terms to the ideal quantum dynamical model. For example, it has been found by numerical simulations in \cite{Marshall_2020,MAS20} that the performance of QAOA may not be monotonically increasing with higher depth when subjected to typical quantum noises. This phenomenon has also been observed in recent experiment \cite{HSM21}. In this case, automatic approach may be able to determine the optimal control depth without using the condition (\ref{intro1}) for complexity control. Secondly, a dataset can be generated for the depth optimization if robust control is considered. For example, by assuming system uncertainties, a large number of samples can be generated for optimizing the control with respect to a single control task \cite{daoyi20,DONG2020242}. Clearly, the algorithm proposed in this paper can be extended to realize the robust depth optimization for QAOA. In this case, cross-validation can be used to determine the critical value $\lambda$ that maximizes the model generalizability and robustness with a train-test split.
	%
	%
	%



	\section*{Acknowledgements}
	This research was supported by the National Natural Science Foundation of China under Grants No. 62173296 and No. 61703364.
	
	\hfill
	

\end{document}